\documentclass[prc,twocolumn,english,showpacs]{revtex4}
\usepackage[T1]{fontenc}
\usepackage[latin1]{inputenc}
\usepackage{babel}
\usepackage{graphics}
\usepackage{multirow}

\makeatletter

\usepackage[T1]{fontenc}
\usepackage[latin1]{inputenc}
\usepackage{babel}
\usepackage{graphics}

\renewcommand {\vec}[1]{\mbox{\boldmath $#1$}}



\makeatother
\begin{document}

\title{Four-body continuum-discretized coupled-channels calculations}

\author{M. Rodr\'{\i}guez-Gallardo$^{1,2,3}$, J.~M.~Arias$^2$,
J.~G\'omez-Camacho$^{2,4}$, A.~M. Moro$^2$,\\ I.~J. Thompson$^{5}$,
and J.~A. Tostevin$^6$}

\affiliation{$^1$ IEM, CSIC, Serrano 123, 28006 Madrid, Spain}

\affiliation{$^2$ Depto.  FAMN, Universidad de Sevilla, Apto. 1065, 41080 Sevilla, Spain}

\affiliation {$^3$ CFNUL, Universidade de Lisboa,
Av. Prof. Gama Pinto 2, 1649-003 Lisboa, Portugal}

\affiliation{$^4$ CNA, Av. Thomas A. Edison, 41092 Sevilla, Spain}

\affiliation{$^5$  LLNL, Physical Science Directorate, P.O. Box 808, L-414, Livermore, CA 94551, USA}

\affiliation{$^6$ Department of Physics, University of Surrey, Guildford, Surrey GU2 7XH, United Kingdom}

\date{\today}

\begin{abstract}
The development of a continuum-bins scheme of discretization for
three-body projectiles is necessary for studies of reactions of
Borromean nuclei such as $^6$He within the continuum-discretized
coupled-channels approach. Such a procedure, for constructing bin
states on selected continuum energy intervals, is formulated and
applied for the first time to reactions of a three-body projectile.
The continuum representation uses the eigenchannel expansion of the
three-body S-matrix. The method is applied to the challenging case
of the $^6$He + $^{208}$Pb reaction at 22 MeV, where an accurate
treatment of both the Coulomb and the nuclear interactions with the
target is necessary.

\end{abstract}
\maketitle

Rapid experimental developments have made studies of the scattering
and reactions of very weakly-bound systems possible in many branches
of physics. Among these are the Efimov states observed in ultra-cold
cesium trimers \cite{Krae06} and molecular triatomic systems, such
as LiHe$_2$ \cite{Bacc00}, halo nuclei \cite{Jen04}, systems
undergoing two-proton radioactivity \cite{Muhk06}, and also
Bose-Einstein condensates and ultra-cold dilute gas studies. The
need for novel and quantitative theoretical descriptions of the
dynamics of such systems, which must include {\it a priori} a realistic
treatment of their continuum spectra, is common to many problems.
Here we discuss continuum effects in the nuclear physics context of
reactions of weakly-bound halo nuclei \cite{Rii94,Han95,Jen04}.
The study of these systems is of special importance in nuclear physics
in order to learn properties of the nucleon-nucleon interaction in 
the low density limit and  investigate the role of pairing off the 
stability line. Many of these exotic nuclei play a significant role
in astrophysical processes. 
Although the method is general, in this Rapid Communication, we will discuss 
reactions of the Borromean halo
nucleus $^6$He, i.e. the bound three-body ($^4$He+n+n) system with
(i) no bound excited states and (ii) no bound binary sub-systems.

Due to their weak binding, halo nuclei are readily excited to the
continuum (broken-up) by the differential forces exerted on the
constituents through 
the nuclear and Coulomb interactions with a target.
Theoretically, the explicit treatment of these breakup channels is
difficult since the states involved are infinite in number and are
not square-integrable. Hence a robust discretization
and truncation scheme, to provide a finite and normalizable basis
to represent the continuum, is required. The continuum-discretized
coupled-channels (CDCC) method, as formulated for two-body
projectiles \cite{Yah86,Aus87}, uses bin states to represent the
continuum. Here, the continuum spectrum is truncated at a maximum
excitation energy $\varepsilon_{\rm max}$ and the model space, from
the breakup threshold to $\varepsilon_{\rm max}$, is then divided
into intervals, where the number and positions of the intervals
can depend on the properties (e.g.~resonant or non-resonant) of the
continuum of the system. For each such interval, or energy bin, a
representative square-integrable state is constructed as a linear
superposition of the two-body scattering states in the interval. The
method has been enormously successful in the description of elastic
and breakup observables in reactions involving weakly-bound two-body
projectiles \cite{Sak86,Nun99,Tos01} and has been recently 
extended to include core excitation \cite{Sum06}.

Of interest here are reactions of three-body projectiles. To date,
published discretized three-body projectile calculations have used a
pseudostate (PS) basis \cite{Mat04,Mat06,Rod08,Ega09}, 
accounting reasonably 
well for existing elastic scattering data.
Based on our own experience \cite{Rod08} of calculations using
PS bases, convergence problems were 
found for reactions where the Coulomb 
interaction plays an important role. 
The alternative continuum treatment using  the energy 
bin technique within the CDCC
has yet to be formulated
and the present work provides this theoretical development.

To describe the three-body ground and excited continuum states of
the projectile we make use of a hyperspherical harmonics (HH)
expansion basis \cite{Dan04}. This involves use of the one radial
and five angular hyperspherical coordinates, $\rho, \alpha,
\widehat{x}, \widehat{y}$, obtained from the normalized Jacobi 
coordinates $\vec{x},$ $\vec{y}$ of the three bodies \cite{Dan04,Rod08}. 
The quantum number
set, $\beta$, that defines each three-body channel \cite{Rod08} are
the hypermomentum $K$, the orbital angular momenta $l_x$ and $l_y$
in coordinates $\vec{x}$ and $\vec{y}$, their total $\vec{l} =
\vec{l}_x +\vec{l}_y$, the total spin $S_x$ of the particles
associated with coordinate $\vec{x}$, and the intermediate summed
angular momentum $\vec{j}_{ab}= \vec{l} +\vec{S}_x$. If the spin of
the third particle, $I$, is assumed fixed then the total angular
momentum is $\vec{j}= \vec{j}_{ab} +\vec{I}$ with projections $\mu$.
Note that, for each continuum energy $\varepsilon$ and total angular momentum $j$,
there will be as many independent solutions
of the three-body scattering problem, as the number
of outgoing channels $\beta$ considered.
These solutions can be chosen as 
the incoming channels $\beta'$,
but any orthogonal combination of these could be 
equally valid.

Based on these total angular momentum eigenstates, ${\mathcal
Y}_{\beta j\mu}(\Omega)$ \cite{Rod08}, where $\Omega \equiv (\alpha,
\widehat{x}, \widehat{y})$, the bin wave functions are defined as
\begin{equation}
\phi^{{\rm bin}}_{n j \mu}(\vec{x},\vec{y})= \sum_{\beta} R^{{\rm bin}}_{n \beta
j}(\rho) {\mathcal Y}_{\beta j \mu}(\Omega), \label{bin}
\end{equation}
where the label $n$ includes reference to the energy
interval of the bin [$\kappa_1,\kappa_2$],  
as well as to the set
of quantum numbers $\beta'$.
The functions $R^{{\rm bin}}_{n \beta j}(\rho)$ in Eq.~(\ref{bin}) 
are the associated hyperradial wave functions,
\begin{eqnarray}  
R^{{\rm bin}}_{n\beta j}(\rho)&\equiv&R^{{\rm bin}}_{[\kappa_1,\kappa_2]\beta'\beta j}(\rho)\label{rbin}\\
&=&\frac{2}{\sqrt{\pi N_{\beta' j}}}\int_{\kappa_1}^{\kappa_2} d\kappa~e^{-i\delta_{\beta' j}(\kappa)}f_{\beta'j}(\kappa) R_{\beta \beta'j}(\kappa,\rho),\nonumber
\end{eqnarray}
where $\kappa=\sqrt{2m|\varepsilon|}/\hbar$ is the momentum associated
to the continuum energy $\varepsilon$, 
$R_{\beta \beta'j}(\kappa\rho)$ are the continuum hyperradial wave functions
with $\delta_{\beta' j}(\kappa)$ their scattering phase shift,
and $f_{\beta'j}(\kappa)$ is a weight function
with $N_{\beta' j}$ its normalization constant.
Note that for each $j$
and three-body energy bin we must construct wave functions for all
allowed incoming channels $\beta'$. Further, as is explicit in Eq.\ (\ref{bin}), for
each $n$ we must also construct $R^{{\rm bin}}_{n \beta j}(\rho)$ for all
allowed outgoing channels $\beta$.

It follows that to include a large number of $\beta'$ channels is a
severe computational challenge, and that it is desirable to
establish a hierarchy of the continuum states according to their
importance to the reaction dynamics. In so doing, we may be able to
describe scattering observables using only a selected set of states,
the number of them depending on the reaction under study.
To this end, we make use of the eigenstates of the multi-channel
three-body S-matrix~\cite{Des06}, or eigenchannels (EC), as follows. 
(i) For each
$j$ and continuum energy $\varepsilon$, the S-matrix in the $\beta$
basis is diagonalized to obtain its EC, enumerated by
$\gamma$, their corresponding eigenvalues $\exp[2i \delta_{\gamma
j}(\kappa)]$ and eigenphases $\delta_{\gamma j}(\kappa)$. (ii) The
magnitudes of these eigenphases are used to order the EC. We will
show that those EC with largest phase shifts are the most strongly
coupled in the reaction dynamics, and thus a hierarchy of states can
be established by such an ordering. This leads to the possibility of
a truncation in the number $n_{ec}$ of EC included and testing of 
the convergence with respect to this number.

Here we apply this methodology to the scattering of $^6$He, treated
as a three-body system of an inert $\alpha$ particle core and two
valence neutrons. A notable property of $^6$He is that none of its
binary sub-systems bind, while the three-body system has a single
bound state with binding energy of $0.973$ MeV and total angular
momentum $j^{\pi}=0^+$. Its low-lying continuum spectrum is
dominated by a narrow $j^{\pi}=2^+$ resonance, $0.825$ MeV above
threshold. We describe the three-body $\alpha$+n+n system with the
same structure model as used in Ref.\ \cite{Rod08}. 
The three-body Hamiltonian includes two-body interactions
plus an effective three-body potential. For a given value of 
the maximum hypermomentum used, $K_\mathrm{max}$, the parameters of 
the latter are adjusted to reproduce the
ground state separation energy and matter radius (for $j=0^+$) 
and the resonance energy (for $j=2^+$) \cite{Rod08}. 
Calculations
were performed using the codes {\sc face} \cite{face} and {\sc
sturmxx} \cite{sturm}. The maximum hypermomentum in the CDCC reaction
calculation was $K_\mathrm{max}
= 8$, which provides converged results for the elastic scattering of 
$^6$He from a heavy target \cite{Rod08}.
The number of channels $\beta$ was
15 ($0^+$), 26 ($1^-$) and 46 ($2^+$). The calculated $^6$He ground
state has binding energy of $0.953$ MeV and a single-particle
density with root mean squared (rms) radius $2.46$ fm, assuming an
$\alpha$ particle rms radius of $1.47$ fm.

\begin{figure}\resizebox*{0.45\textwidth}{!}{\includegraphics{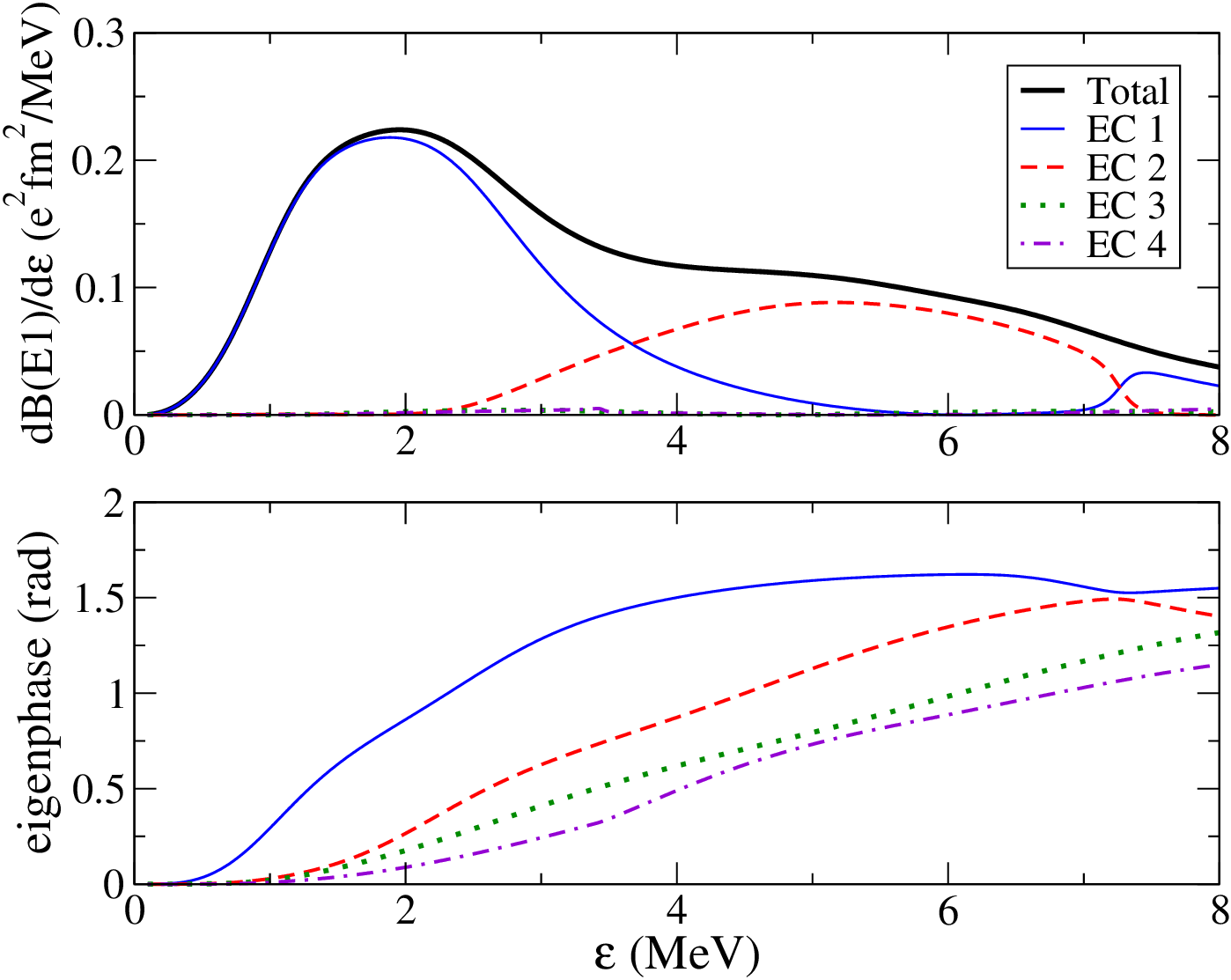}}
\vspace{-0.4cm} \caption{\label{be1} (Color online) $B(E1)$ distribution (upper
panel) and eigenphase shifts (lower panel) for the first few $j^{\pi}=1^-$ EC of
$^6$He. The thick solid line in the upper panel is the total 
$B(E1)$ distribution.}
\end{figure}

\begin{figure}\resizebox*{0.45\textwidth}{!}{\includegraphics{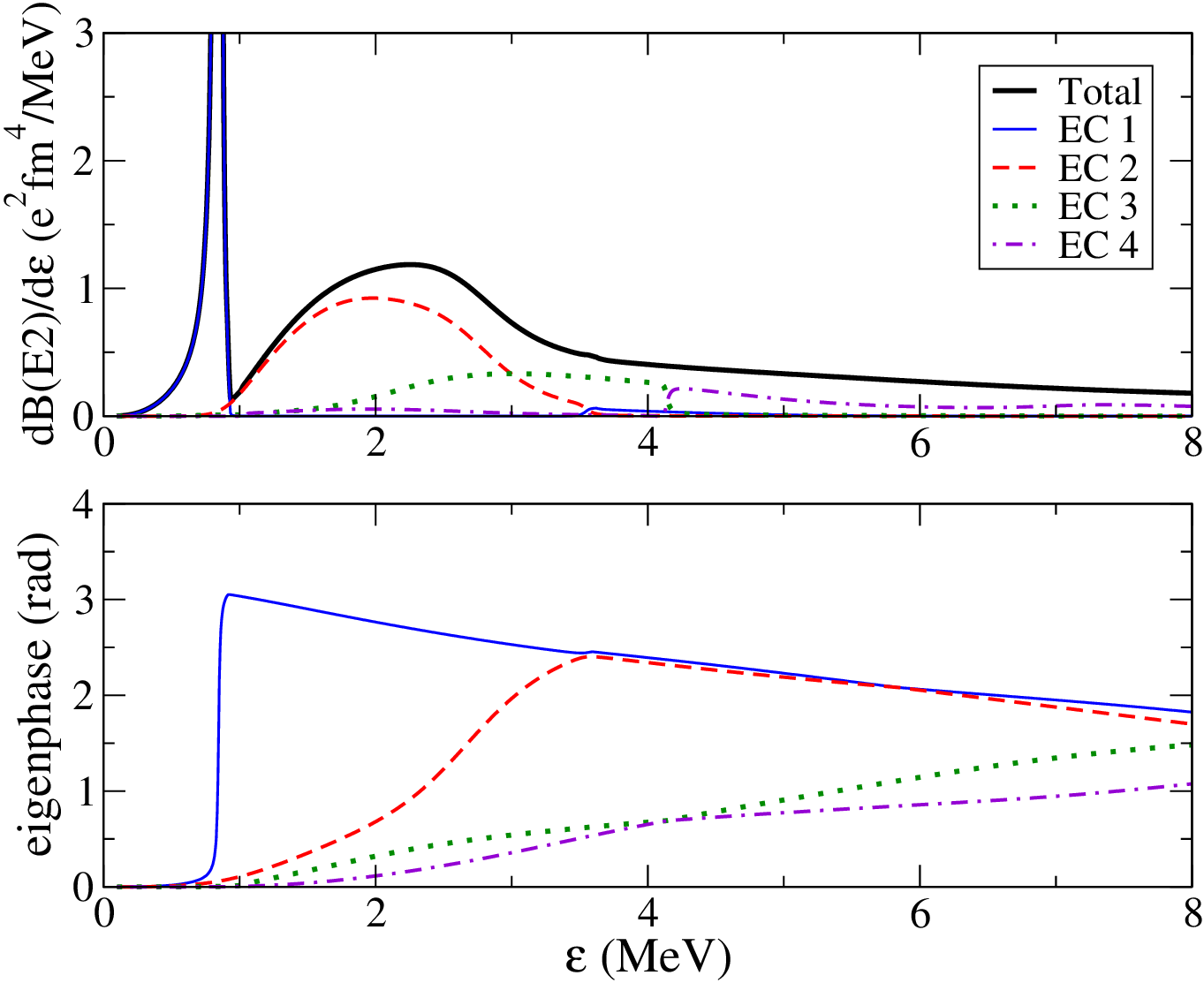}}
\vspace{-0.4cm} \caption{\label{be2} (Color online) $B(E2)$ distribution (upper
panel) and eigenphase shifts (lower panel) for the first few $j^{\pi}=2^+$ EC of
$^6$He. The thick solid line in the upper panel is the total 
$B(E2)$ distribution.}
\end{figure}

The EC basis states have properties that are useful to describe
collisions. The upper panels of Figs.\ \ref{be1} and \ref{be2} show
the Coulomb $B(E1)$ and $B(E2)$ transition probabilities,
respectively, as a function of the $^6$He excitation energy above
threshold. The curves show the total $B(E1)$ and $B(E2)$ strengths
(thick solid) and the contributions from the first four EC. Since
the $^6$He(g.s.) has $j^{\pi}=0^+$, $B(E1)$ measures the electric
dipole strength to $1^-$ states, and $B(E2)$ the quadrupole strength
to $2^+$ states. For each EC, the $1^-$ and $2^+$ eigenphases
are also shown in the lower panels of the figures.
It has been shown (see for instance Ref. \cite{Dan98}) that the 
position of the peak of  
 the  $B(E1)$ distribution depends on the maximum hypermomentum 
chosen. 
In particular, for  $K_\mathrm{max}=10$  the maximum appears 
around 2 MeV, and decreases to about 1 MeV, for 
$K_\mathrm{max}=20$, for which convergence of 
this observable is achieved. For  $K_\mathrm{max}=8$, which is the value 
used in this work, the maximum is around 2 MeV, and hence this observable is 
not converged.  Nevertheless,  
as we have shown 
in Ref.~\cite{Rod08}, the scattering calculations are converged 
with this value, provided that  the parameters of the three-body potential
are adjusted such that the $^6$He system has the same binding energy 
and rms matter radius for the chosen $K_\mathrm{max}$.
In principle, the reaction calculations could be done
with a higher  $K_\mathrm{max}$ value, although this would turn
these calculations computationally very demanding. 



It is evident from Figs.~\ \ref{be1} and \ref{be2}
that there is energy localization for the EC
contributions to the $E\lambda$ strength. Also evident 
is that those
EC with eigenphases of the largest magnitude make the maximum
contribution to the $B(E\lambda)$ at low excitation energy. 
This is
a very appealing feature for scattering studies where, at moderate
collision energies, only continuum states up to a few MeV play a
significant role in the reaction process. So, for example, if
excitation energies up to $\approx8$ MeV are strongly populated in
a particular reaction, the first three or four EC should suffice to
describe the relevant part of the continuum. Of course, in nuclear
collisions near and above the Coulomb barrier, nuclear forces (not
included in Figs.\ \ref{be1} and \ref{be2}) will play an important
role. We show below that, with nuclear interactions present, only a
few EC are needed for converged scattering observables.

It is also interesting to identify the dominant quantum numbers of
the EC with the largest contributions to these observables. 
For $B(E1)$ the most important component of EC 1 and 2 
has $l_{nn}=l_x=0,S_x=0$, or a {\em dineutron} configuration. 
This dominance, could explain that previous two-body
($\alpha$+2n) model calculations of $^6$He reactions work 
to a limited extent, although it is well known that such simple models
overestimate continuum coupling effects without
parameter adjustments \cite{Rus05,Mor07}.
For $B(E2)$ the most relevant structure is the $2^+$ resonance 
that corresponds to the EC 1. In this case we have  a
combination of three different configurations: a dineutron ($l_x=0,
S_x=0$), a cigar-like structure ($l_x=2, S_x=0$), and a spin-triplet
structure ($l_x=1, S_x=1$).


We now compute the states of Eq.\ (\ref{bin})
and use these as the basis states for a four-body CDCC calculation
using the techniques of Ref.~\cite{Rod08}. There, a multipole
expansion of the coupling potentials was developed for a three-body
projectile plus target system. The weight function $f_{\gamma j}(\kappa)$
(with $\gamma$ instead of $\beta'$) 
used in the superposition of states when constructing the bins 
Eq.~(\ref{rbin}) was taken to be
unity for the non-resonant ($0^+$, $1^-$, and $2^+$) continuum and 
$\sin{(\delta_{\gamma j}(\kappa))}$ for the resonant ($2^+$) continuum. The
latter provides an improved description of the resonant character.

Four-body CDCC calculations are carried out for the $^6$He +
$^{208}$Pb reaction at 22 MeV, for which experimental data are
available from Ref.~\cite{San07}. This
is one of the more demanding examples, due to (a) the importance of
long-range Coulomb couplings arising from the heavy target and (b)
the incident energy being near the Coulomb barrier and Coulomb and
nuclear forces both playing a significant role.

Implicit in the CDCC is that the actual continuum can be truncated
at a maximum excitation energy $\varepsilon_{\rm max}$, whose choice
is dependent on the specific reaction and projectile energy. The
convergence of the model space calculation must then be studied with
respect to $\varepsilon_{\rm max}$ and the number of bins assumed 
($n_{\rm bin}$).
For our four-body case we have, in addition, to study convergence
with respect to $n_{ec}$, the number of EC included. The present
calculations include $^6$He states with $j^\pi=0^+,1^-$ and $2^+$
and projectile-target interaction multipole couplings with order
$Q=0,1,2$. Coulomb and nuclear potentials are included. The nuclear
interactions with the target used parametrized optical potentials.
For $n+^{208}$Pb and $\alpha+^{208}$Pb the potentials were from
Refs.\ \cite{Rob91} and \cite{Bar74}, respectively.

The coupled
equations describing the projectile-target motion were solved with
the code {\sc fresco} \cite{Thom88}, the coupling form factors being
read from external files. 
Partial waves up to $J=150$ were included and the solutions
were matched to their asymptotic
forms at radius $R_m=200$ fm.  
In order to check the convergence with respect to the basis size,
for the different observables shown in this work,
we have used five different sets that are summarized in Table~\ref{tab}.
Sets I, II, and III have the same maximum energy $\varepsilon_{\rm max}$ 
and number of EC $n_{\rm ec}$, so they are used to test the convergence
with respect to the number of bins considered for each EC and $j^{\pi}$.
Set IV is like set II but including the fifth EC, so it will be used to
study the convergence with $n_{\rm ec}$. Finally, set V  is like set II
but including an additional bin with $\varepsilon$=8-9 MeV for all EC and 
$j^{\pi}$, providing information on the convergence with respect to 
$\varepsilon_{\rm max}$.

\begin{table}[tb]
\begin{tabular}{lcccccc} 
\hline
Set &  $\varepsilon_{\rm max}$ (MeV)  & $n_{\rm ec}$  & $n_{\rm bin}(0^+)$  & $n_{\rm bin}(1^-)$ & $n_{\rm bin}(2^+)$ & $N$ \\
\hline
I & 8 & 4 & 6 & 9 &6 & 85 \\
II & 8 & 4 & 9 & 12 &9 & 121 \\ 
III & 8 & 4 & 12 & 15 & 12 & 157\\
IV & 8 & 5 (1-4,5) & (9,6) & (12,9) & (9,6) & 142 \\
V & 9 (0-8,8-9) & 4 & (9,1) & (12,1) & (9,1) & 133\\   
\hline
\end{tabular}
\caption{Different sets of states used in this work for the four-body 
CDCC calculations.
$N$ is the total number of states considered (including the ground state).
The parentheses are used to specify  the two different 
bin schemes adopted depending on the EC (Set IV) or on the energy (Set V).
\label{tab}}
\end{table}

\begin{figure}\resizebox*{0.45\textwidth}{!}{\includegraphics{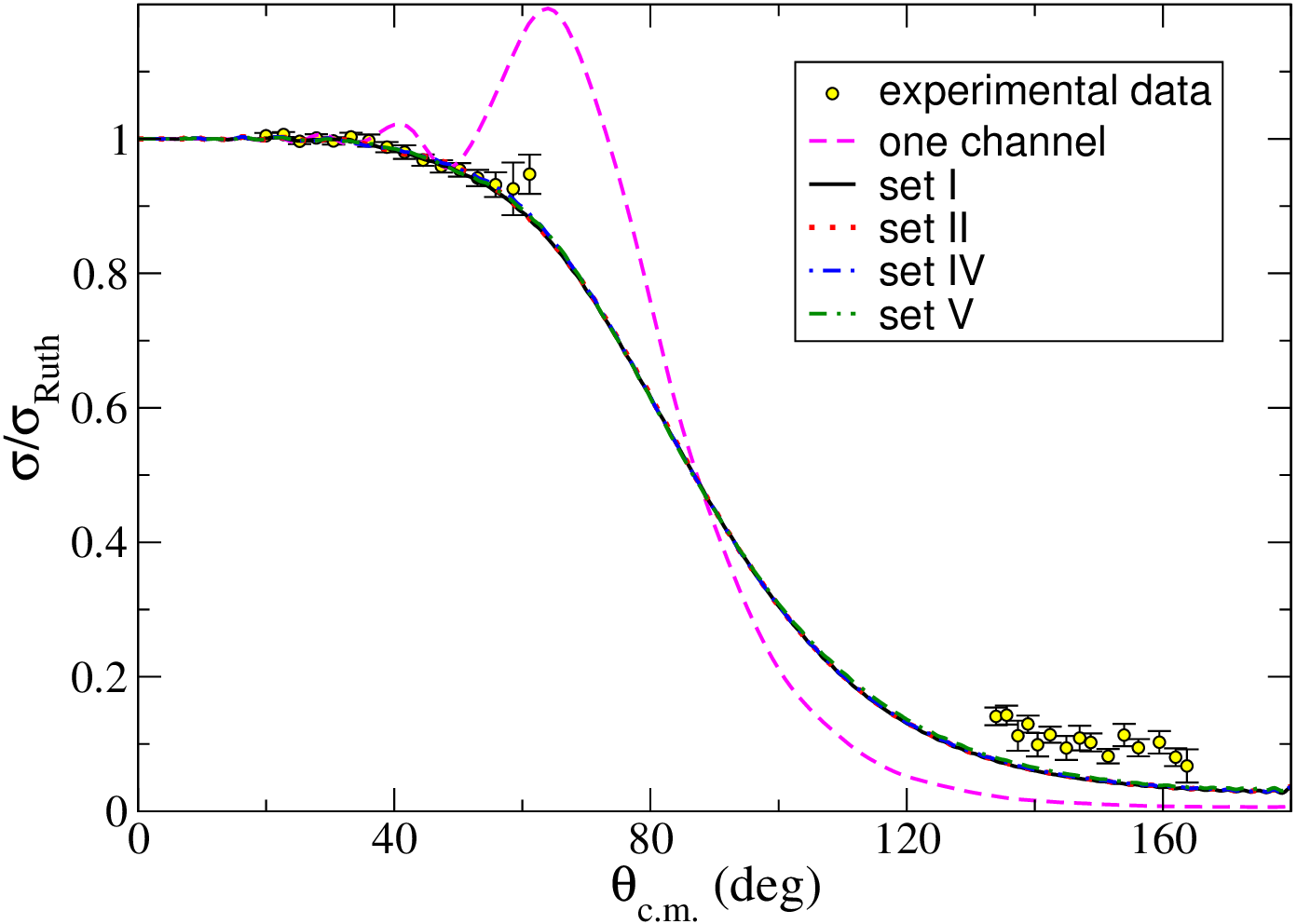}}
\vspace{-0.4cm} \caption{\label{he6pb_el} (Color online) Elastic
differential cross section (ratio to Rutherford) in the center of
mass frame for the $^6$He$+^{208}$Pb reaction at 22 MeV. See text for details.
The experimental data are taken from Ref.~\cite{San07}.
}
\end{figure}

First, we analyze the elastic scattering.
Fig.\ \ref{he6pb_el} compares the calculated and experimental
(circles) elastic differential cross section angular distributions
(ratio to Rutherford). The dashed line is the one channel
calculation, in which continuum states are omitted. The solid
(set I), the dotted (set II), dot-dashed (set IV) and  
double dot-dashed (set V) lines
are the full CDCC calculations with different choices of the model space.
It is remarkable that all these calculations
are almost indistinguishable,
showing that the elastic angular distribution is 
converged  with set I. 
Also, this full four-body CDCC calculation, 
in contrast to the one channel calculation, 
describes the data fairly well.
Set III is not shown since the convergence with 
the number of bins is already seen with sets I and II.

\begin{figure}\resizebox*{0.45\textwidth}{!}{\includegraphics{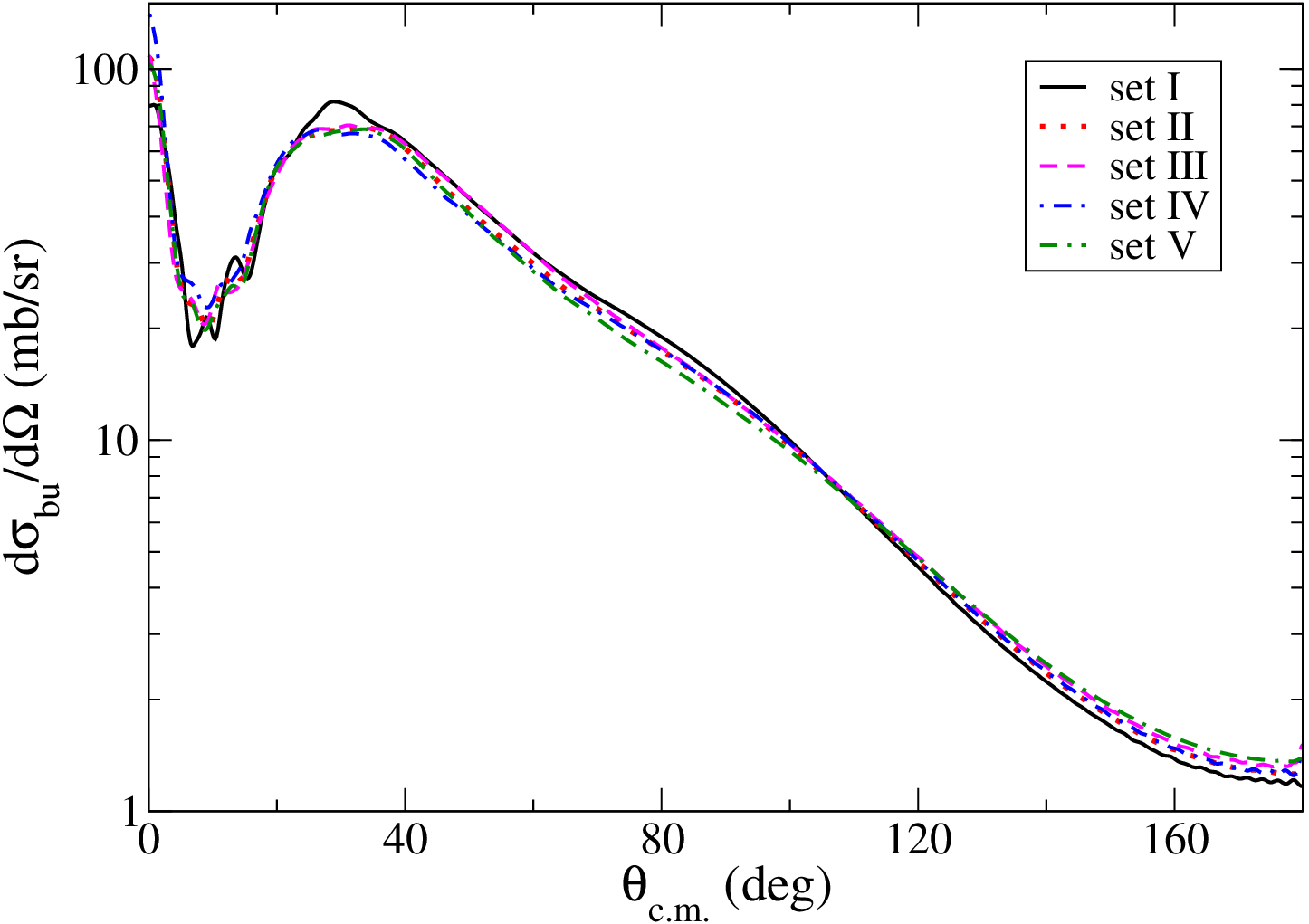}}
\caption{\label{he6pb_bu} (Color online) Breakup differential cross section 
angular distribution  in the center of mass frame for the $^6$He$+^{208}$Pb 
reaction at 22 MeV. 
See text for details.
}
\end{figure}

Solution of the CDCC coupled equations for the projectile-target
S-matrix also provides predictions for breakup observables. 
First, Fig.\ \ref{he6pb_bu} shows the breakup differential cross
section  angular distribution,  summed over the excitation energy. 
The solid
(set I), dotted (set II), dashed (set III), dot-dashed (set IV) and  
double dot-dashed (set V) lines
are the full CDCC calculations with different choices of the model space,
as in Fig\ \ref{he6pb_el}.
For this breakup observable the rate of convergence is slightly slower than
for the elastic cross section, although the convergence reached with set II
is fairly good.

\begin{figure*}\resizebox*{0.70\textwidth}{!}{\includegraphics{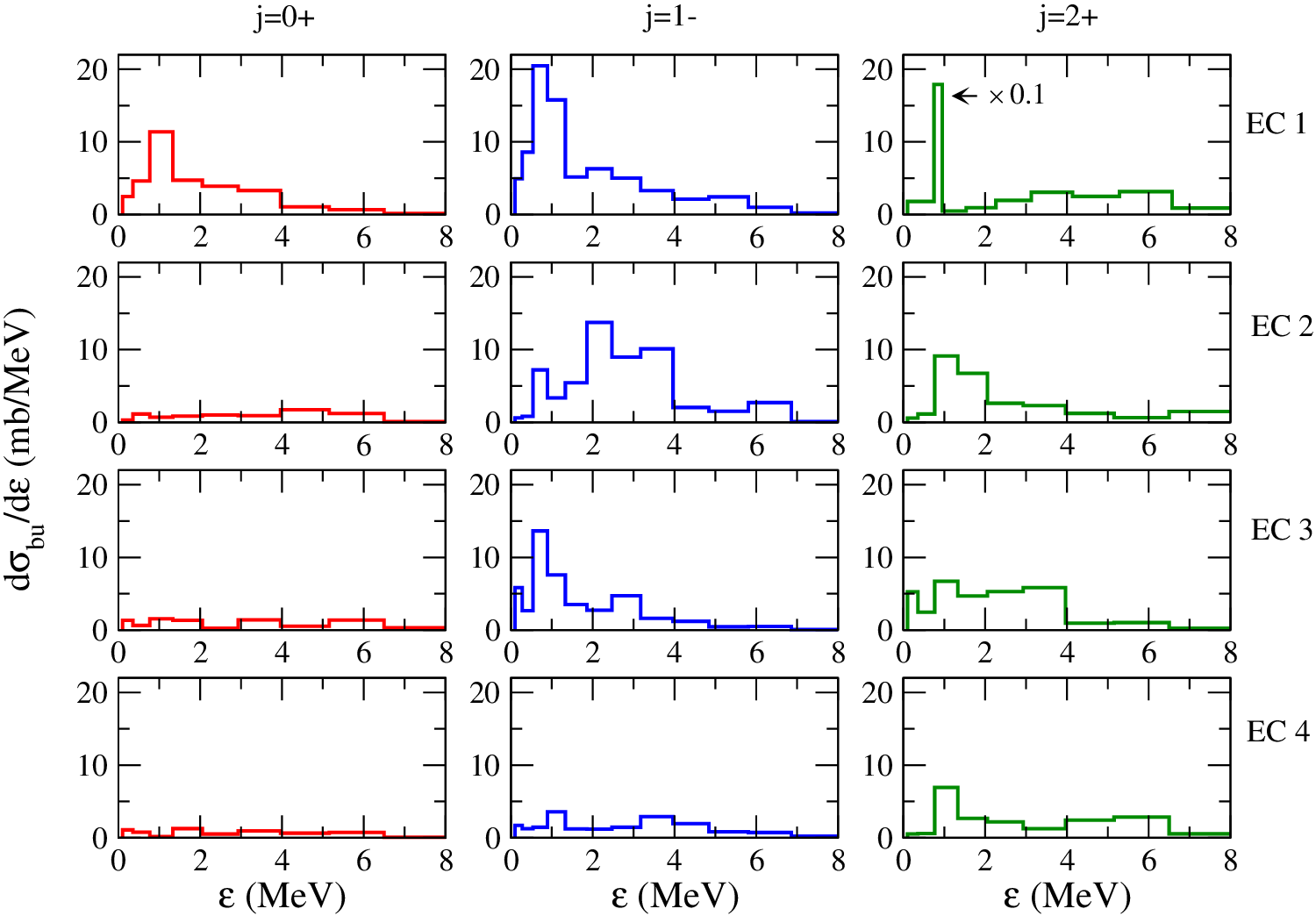}}
\caption{\label{buvsEx} (Color online) Breakup cross section distribution
for the $^6$He$+^{208}$Pb reaction at 22
MeV with respect
to $^6$He excitation energy (over the breakup threshold) for each $j^\pi$ 
and EC included in the calculation.  Basis set II was used.}
\end{figure*}

Second, Fig.\ \ref{buvsEx} shows the breakup cross section distribution, 
as a function of $^6$He excitation energy above breakup threshold, for each
$j^{\pi}$ and for each EC. 
This distribution is calculated by dividing the breakup
cross section to each bin by its energy width.
We use here set II that gives a converged result 
for the breakup angular distribution. 
As was anticipated, in all $j^{\pi}$,
the contribution of each EC decreases with decreasing magnitude of
its eigenphase.
In particular, the contribution of
the fourth EC is already small, justifying the use of a truncated
calculation and $n_{ec}\approx 4$. We have also checked that
the contribution of the fifth EC (comparing sets II and IV) 
changes the total breakup cross section by less than 2\%.
This figure also shows the importance of the dipole and resonant
states. These give the largest contribution to the breakup cross section.


Collisions of loosely bound three-body projectiles with a target
can be studied in the continuum-discretized coupled-channels
framework. The continuum-bin discretization procedure has been
formulated and used for the first time for a three-body projectile,
requiring the superposition of three-body scattering states. For
each three-body total angular momentum $j$ and energy $\varepsilon$,
scattering states are calculated for ingoing boundary conditions
with quantum numbers $\beta$. The multi-channel S-matrix in this
$\beta$ basis, determined from the asymptotics of these states, is
diagonalized to determine the eigenchannels (EC) that are ordered
according to the magnitude of their associated eigenphases. We have
shown that those EC with the largest eigenphases are most strongly
coupled in the collision, suggesting a hierarchy of continuum states
and an associated truncation scheme.
This EC hierarchy
allows practical CDCC calculations that
extend the  energy binning procedure to four-body reactions induced
by three-body projectiles. The formalism has been applied to the $^6$He+
$^{208}$Pb reaction at 22 MeV.
The convergence of the results for 
elastic and breakup cross sections have been presented. 
Good agreement with existing elastic data for this reaction has been obtained.

\begin{acknowledgments}
We are grateful to F.M.\ Nunes and R.C.\ Johnson
for useful discussions
and suggestions.
This work was supported in part by the FCT under the Grant
POCTI/ISFL/2/275
and in part by the DGICYT
under Projects FIS 2008-04189, FPA 2006-13807-C02-01, and 
consolider CPAN. 
Part of this work
was performed under the auspices of the U.S. Department of Energy by
Lawrence Livermore National Laboratory under Contract
DE-AC52-07NA27344. J.A.T. acknowledges the support of the the United
Kingdom Science and Technology Facilities Council (STFC) under Grant
EP/D003628. 
A.M.M. acknowledges a research grant from the Junta de Andaluc\'ia. 
\end{acknowledgments}


\bibliographystyle{apsrev}
\bibliography{./4bCDCC}

\begin{thebibliography}{27}
\expandafter\ifx\csname natexlab\endcsname\relax\def\natexlab#1{#1}\fi
\expandafter\ifx\csname bibnamefont\endcsname\relax
  \def\bibnamefont#1{#1}\fi
\expandafter\ifx\csname bibfnamefont\endcsname\relax
  \def\bibfnamefont#1{#1}\fi
\expandafter\ifx\csname citenamefont\endcsname\relax
  \def\citenamefont#1{#1}\fi
\expandafter\ifx\csname url\endcsname\relax
  \def\url#1{\texttt{#1}}\fi
\expandafter\ifx\csname urlprefix\endcsname\relax\def\urlprefix{URL }\fi
\providecommand{\bibinfo}[2]{#2}
\providecommand{\eprint}[2][]{\url{#2}}

\bibitem[{\citenamefont{{T.~Kraemer~{\it et al.}}}(2006)}]{Krae06}
\bibinfo{author}{\bibnamefont{{T.~Kraemer~{\it et al.}}}},
  \bibinfo{journal}{Nature} \textbf{\bibinfo{volume}{440}},
  \bibinfo{pages}{315} (\bibinfo{year}{2006}).

\bibitem[{\citenamefont{{I.~Baccarelli~{\it et al.}}}(2000)}]{Bacc00}
\bibinfo{author}{\bibnamefont{{I.~Baccarelli~{\it et al.}}}},
  \bibinfo{journal}{Phys. Chem. Chem. Phys.} \textbf{\bibinfo{volume}{2}},
  \bibinfo{pages}{4067} (\bibinfo{year}{2000}).

\bibitem[{\citenamefont{{A.~S.~Jensen~{\it et al.}}}(2004)}]{Jen04}
\bibinfo{author}{\bibnamefont{{A.~S.~Jensen~{\it et al.}}}},
  \bibinfo{journal}{Rev. Mod. Phys.} \textbf{\bibinfo{volume}{76}},
  \bibinfo{pages}{215} (\bibinfo{year}{2004}).

\bibitem[{\citenamefont{{I.~Muhka~{\it et al.}}}(2006)}]{Muhk06}
\bibinfo{author}{\bibnamefont{{I.~Muhka~{\it et al.}}}},
  \bibinfo{journal}{Nature} \textbf{\bibinfo{volume}{439}},
  \bibinfo{pages}{298} (\bibinfo{year}{2006}).

\bibitem[{\citenamefont{Riisager}(1994)}]{Rii94}
\bibinfo{author}{\bibfnamefont{K.}~\bibnamefont{Riisager}},
  \bibinfo{journal}{Rev. Mod. Phys.} \textbf{\bibinfo{volume}{66}},
  \bibinfo{pages}{1105} (\bibinfo{year}{1994}).

\bibitem[{\citenamefont{Hansen et~al.}(1995)\citenamefont{Hansen, Jensen, and
  Jonson}}]{Han95}
\bibinfo{author}{\bibfnamefont{P.~G.} \bibnamefont{Hansen}},
  \bibinfo{author}{\bibfnamefont{A.~S.} \bibnamefont{Jensen}},
  \bibnamefont{and} \bibinfo{author}{\bibfnamefont{B.}~\bibnamefont{Jonson}},
  \bibinfo{journal}{Ann. Rev. Nucl. Part. Sci.} \textbf{\bibinfo{volume}{45}},
  \bibinfo{pages}{591} (\bibinfo{year}{1995}).

\bibitem[{\citenamefont{{M.~Yahiro~{\it et al.}}}(1986)}]{Yah86}
\bibinfo{author}{\bibnamefont{{M.~Yahiro~{\it et al.}}}},
  \bibinfo{journal}{Prog.\ Theor.\ Phys.\ Suppl.}
  \textbf{\bibinfo{volume}{89}}, \bibinfo{pages}{32} (\bibinfo{year}{1986}).

\bibitem[{\citenamefont{{N.~Austern~{\it et al.}}}(1987)}]{Aus87}
\bibinfo{author}{\bibnamefont{{N.~Austern~{\it et al.}}}},
  \bibinfo{journal}{Phys. Rep.} \textbf{\bibinfo{volume}{154}},
  \bibinfo{pages}{125} (\bibinfo{year}{1987}).

\bibitem[{\citenamefont{Sakuragi et~al.}(1986)\citenamefont{Sakuragi, Yahiro,
  and Kamimura}}]{Sak86}
\bibinfo{author}{\bibfnamefont{Y.}~\bibnamefont{Sakuragi}},
  \bibinfo{author}{\bibfnamefont{M.}~\bibnamefont{Yahiro}}, \bibnamefont{and}
  \bibinfo{author}{\bibfnamefont{M.}~\bibnamefont{Kamimura}},
  \bibinfo{journal}{Prog. Theor. Phys. Suppl.} \textbf{\bibinfo{volume}{89}},
  \bibinfo{pages}{136} (\bibinfo{year}{1986}).

\bibitem[{\citenamefont{Nunes and Thompson}(1999)}]{Nun99}
\bibinfo{author}{\bibfnamefont{F.~M.} \bibnamefont{Nunes}} \bibnamefont{and}
  \bibinfo{author}{\bibfnamefont{I.~J.} \bibnamefont{Thompson}},
  \bibinfo{journal}{Phys. Rev. C} \textbf{\bibinfo{volume}{59}},
  \bibinfo{pages}{2652} (\bibinfo{year}{1999}).

\bibitem[{\citenamefont{Tostevin et~al.}(2001)\citenamefont{Tostevin, Nunes,
  and Thompson}}]{Tos01}
\bibinfo{author}{\bibfnamefont{J.~A.} \bibnamefont{Tostevin}},
  \bibinfo{author}{\bibfnamefont{F.~M.} \bibnamefont{Nunes}}, \bibnamefont{and}
  \bibinfo{author}{\bibfnamefont{I.~J.} \bibnamefont{Thompson}},
  \bibinfo{journal}{Phys. Rev. C} \textbf{\bibinfo{volume}{63}},
  \bibinfo{pages}{024617} (\bibinfo{year}{2001}).

\bibitem[{\citenamefont{Summers et~al.}(2006)\citenamefont{Summers, Nunes, and
  Thompson}}]{Sum06}
\bibinfo{author}{\bibfnamefont{N.~C.} \bibnamefont{Summers}},
  \bibinfo{author}{\bibfnamefont{F.~M.} \bibnamefont{Nunes}}, \bibnamefont{and}
  \bibinfo{author}{\bibfnamefont{I.~J.} \bibnamefont{Thompson}},
  \bibinfo{journal}{Phys. Rev. C} \textbf{\bibinfo{volume}{74}},
  \bibinfo{pages}{014606} (\bibinfo{year}{2006}).

\bibitem[{\citenamefont{{T.~Matsumoto~{\it et al.}}}(2004)}]{Mat04}
\bibinfo{author}{\bibnamefont{{T.~Matsumoto~{\it et al.}}}},
  \bibinfo{journal}{Phys. Rev. C} \textbf{\bibinfo{volume}{70}},
  \bibinfo{pages}{061601} (\bibinfo{year}{2004}).

\bibitem[{\citenamefont{{T.~Matsumoto~{\it et al.}}}(2006)}]{Mat06}
\bibinfo{author}{\bibnamefont{{T.~Matsumoto~{\it et al.}}}},
  \bibinfo{journal}{Phys. Rev. C} \textbf{\bibinfo{volume}{73}},
  \bibinfo{pages}{051602} (\bibinfo{year}{2006}).

\bibitem[{\citenamefont{{M.~Rodr\'{\i}guez-Gallardo~{\it et
  al.}}}(2008)}]{Rod08}
\bibinfo{author}{\bibnamefont{{M.~Rodr\'{\i}guez-Gallardo~{\it et al.}}}},
  \bibinfo{journal}{Phys. Rev. C} \textbf{\bibinfo{volume}{77}},
  \bibinfo{pages}{064609} (\bibinfo{year}{2008}).

\bibitem[{\citenamefont{Egami et~al.}(2009)\citenamefont{Egami, Matsumoto,
  Ogata, and Yahiro}}]{Ega09}
\bibinfo{author}{\bibfnamefont{T.}~\bibnamefont{Egami}},
  \bibinfo{author}{\bibfnamefont{T.}~\bibnamefont{Matsumoto}},
  \bibinfo{author}{\bibfnamefont{K.}~\bibnamefont{Ogata}}, \bibnamefont{and}
  \bibinfo{author}{\bibfnamefont{M.}~\bibnamefont{Yahiro}},
  \bibinfo{journal}{Prog. Theor. Phys.} \textbf{\bibinfo{volume}{121}},
  \bibinfo{pages}{789} (\bibinfo{year}{2009}).

\bibitem[{\citenamefont{{B.~V.~Danilin~{\it et al.}}}(2004)}]{Dan04}
\bibinfo{author}{\bibnamefont{{B.~V.~Danilin~{\it et al.}}}},
  \bibinfo{journal}{Phys. Rev. C} \textbf{\bibinfo{volume}{69}},
  \bibinfo{pages}{024609} (\bibinfo{year}{2004}).

\bibitem[{\citenamefont{Descouvemont et~al.}(2006)\citenamefont{Descouvemont,
  Tursunov, and Baye}}]{Des06}
\bibinfo{author}{\bibfnamefont{P.}~\bibnamefont{Descouvemont}},
  \bibinfo{author}{\bibfnamefont{E.}~\bibnamefont{Tursunov}}, \bibnamefont{and}
  \bibinfo{author}{\bibfnamefont{D.}~\bibnamefont{Baye}},
  \bibinfo{journal}{Nucl. Phys.} \textbf{\bibinfo{volume}{A765}},
  \bibinfo{pages}{370} (\bibinfo{year}{2006}).

\bibitem[{\citenamefont{Thompson et~al.}(2004)\citenamefont{Thompson, Nunes,
  and Danilin}}]{face}
\bibinfo{author}{\bibfnamefont{I.~J.} \bibnamefont{Thompson}},
  \bibinfo{author}{\bibfnamefont{F.~M.} \bibnamefont{Nunes}}, \bibnamefont{and}
  \bibinfo{author}{\bibfnamefont{B.~V.} \bibnamefont{Danilin}},
  \bibinfo{journal}{Comput. Phys. Commun.} \textbf{\bibinfo{volume}{161}},
  \bibinfo{pages}{87} (\bibinfo{year}{2004}).

\bibitem[{\citenamefont{Thompson}(2002)}]{sturm}
\bibinfo{author}{\bibfnamefont{I.~J.} \bibnamefont{Thompson}},
  \bibinfo{journal}{Unpublished. Users manual available from the author}
  (\bibinfo{year}{2002}).

\bibitem[{\citenamefont{{B.~V.~Danilin~{\it et al.}}}(1998)}]{Dan98}
\bibinfo{author}{\bibnamefont{{B.~V.~Danilin~{\it et al.}}}},
  \bibinfo{journal}{Nucl. Phys.} \textbf{\bibinfo{volume}{A632}},
  \bibinfo{pages}{383} (\bibinfo{year}{1998}).

\bibitem[{\citenamefont{{K.~Rusek~{\it et al.}}}(2005)}]{Rus05}
\bibinfo{author}{\bibnamefont{{K.~Rusek~{\it et al.}}}},
  \bibinfo{journal}{Phys. Rev. C} \textbf{\bibinfo{volume}{72}},
  \bibinfo{pages}{037603} (\bibinfo{year}{2005}).

\bibitem[{\citenamefont{{A.~M.~Moro~{\it et al.}}}(2007)}]{Mor07}
\bibinfo{author}{\bibnamefont{{A.~M.~Moro~{\it et al.}}}},
  \bibinfo{journal}{Phys. Rev. C} \textbf{\bibinfo{volume}{75}},
  \bibinfo{pages}{064607} (\bibinfo{year}{2007}).

\bibitem[{\citenamefont{{A.~M.~S\'anchez-Ben\'{\i}tez~{\it et
  al.}}}(2008)}]{San07}
\bibinfo{author}{\bibnamefont{{A.~M.~S\'anchez-Ben\'{\i}tez~{\it et al.}}}},
  \bibinfo{journal}{Nucl. Phys.} \textbf{\bibinfo{volume}{A803}},
  \bibinfo{pages}{30} (\bibinfo{year}{2008}).

\bibitem[{\citenamefont{{M.~L.~Roberts~{\it et al.}}}(1991)}]{Rob91}
\bibinfo{author}{\bibnamefont{{M.~L.~Roberts~{\it et al.}}}},
  \bibinfo{journal}{Phys. Rev. C} \textbf{\bibinfo{volume}{44}},
  \bibinfo{pages}{2006} (\bibinfo{year}{1991}).

\bibitem[{\citenamefont{Barnett and Lilley}(1974)}]{Bar74}
\bibinfo{author}{\bibfnamefont{A.~R.} \bibnamefont{Barnett}} \bibnamefont{and}
  \bibinfo{author}{\bibfnamefont{J.~S.} \bibnamefont{Lilley}},
  \bibinfo{journal}{Phys. Rev. C} \textbf{\bibinfo{volume}{9}},
  \bibinfo{pages}{2010} (\bibinfo{year}{1974}).

\bibitem[{\citenamefont{Thompson}(1988)}]{Thom88}
\bibinfo{author}{\bibfnamefont{I.~J.} \bibnamefont{Thompson}},
  \bibinfo{journal}{Comp. Phys. Rep.} \textbf{\bibinfo{volume}{7}},
  \bibinfo{pages}{167} (\bibinfo{year}{1988}).

\end{thebibliography}

\end{document}